\documentclass[draft,aps,floatfix]{revtex4}
\usepackage{graphicx}

\newcommand{\bea}{\begin{eqnarray}}
\newcommand{\eea}{\end{eqnarray}}

\begin{document}

\title{Strong decays of P wave baryons in the $1/N_c$ expansion
\footnote{Contribution to the Proc. of the ``Large $N_c$ QCD 2004" workshop, July 2004, ECT$^*$, Trento, Italy.  } }

\author{N.N. Scoccola}

\affiliation{Physics Depart., Comisi\'on Nacional de Energ\'{\i}a
At\'omica,
     (1429) Buenos Aires, Argentina.\\
CONICET, Rivadavia 1917, (1033) Buenos Aires, Argentina.\\
Universidad Favaloro, Sol{\'\i}s 453, (1078) Buenos Aires,
Argentina.}

\begin{abstract}
The hadronic decays of non-strange negative parity baryons are analyzed in the
framework of the $1/N_c$ expansion. A complete basis of spin-flavor operators for the
pseudoscalar meson $S$ and $D$ partial wave amplitudes is established to order $1/N_c$
and the unknown effective coefficients are determined by fitting to the empirically
known widths. A  set of relations between widths that result at the leading order,
i.e. order $N_c^0$, is given and tested with the available data.
\end{abstract}

\maketitle

\section{Introduction}

The  $1/N_c$ expansion has proven  to be a very useful tool for
analyzing the baryon sector. This success  is mostly a consequence
of the emergent contracted spin-flavor symmetry in the large $N_c$
limit\cite{GervaisSakita,DashenManohar}. For ground state baryons
(identified for $N_c=3$ with the spin 1/2 octet and spin 3/2 decuplet
in the case of three flavors), that symmetry gives rise to several important
relations that hold at different orders in the  $1/N_c$
expansion\cite{GroundStates1,Georgi,GroundStates2}. The domain of excited
baryons (baryon resonances) has also been explored in the framework
of the $1/N_c$ expansion\cite{Goity}$^-$\cite{PirjolSchat}
with very promising results. The analyses carried out so far have
been constrained to states that belong to a definite spin-flavor
and orbital multiplet, i.e. the possibility of mixing of different
such multiplets (so called configuration mixing) has been
disregarded. It is very likely that such effects are small for dynamical
reasons. Indeed, it has been shown that the only configuration mixings
that are not suppressed by $1/N_c$ factors involve couplings to the orbital
degrees of freedom\cite{GoityMixDecay}. The $1/N_c$ analyses of excited
baryon masses have shown that orbital angular momentum couplings
turn out to be very small\cite{CCGL,GSS}, which is in agreement
with older results in the quark model\cite{CapstickRoberts}. This
strongly suggests that a similar suppression, which is not a
consequence of the $1/N_c$ expansion but rather of QCD dynamics,
also takes place in configuration mixings. Thus, disregarding
configuration mixing is  likely to be a good approximation for the
purpose of phenomenology.  Within such a framework, a few analyses
of excited baryon strong decays have been carried out, namely the decays
of the negative parity SU(6) 70-plet\cite{CaroneGeorgi} and of
the Roper 56-plet\cite{CarlsonCarone}. In the
case of interest in the present work, namely the 70-plet, the analysis in
Ref.~\cite{CaroneGeorgi}  used an incomplete basis of
operators at  sub-leading order  in $1/N_c$.   The main
motivation of the present work (see Ref.~\cite{GSS05} for more details)
is to provide a general framework for the study of the strong decays and
to perform a complete analysis to ${\it{O}}(1/N_c)$ in the particular case of the
decays of the
non-strange members of the 70-plet (i.e., the mixed symmetry  20-plet of SU(4)) into
ground state baryons plus a pion or an eta meson.

This contribution is organized as follows: Section 2 contains the framework for calculating
the decays, Section 3 provides the basis of effective operators, Section 4 presents
the results, and finally the conclusions are given in Section~5.

\section{ Framework for decays  }

As discussed  above in the present application of the $1/N_c$ expansion to excited baryons the assumption is made
that there is an approximate spin-flavor symmetry. Thus,
the excited baryons are therefore classified in multiplets of the $O(3)\times SU(2N_f)$ group. $O(3)$
corresponds to spatial rotations and $SU(2N_f)$ is the spin-flavor group where $N_f$ is the number of flavors
being considered, equal to two in the present work. The ground state baryons, namely the $N$ and $\Delta$ states,
belong to the $({\bf 1},{\bf 20_S})$ representation, where the $\bf 20_S$ is the  totally symmetric
representation of $SU(4)$. The
negative parity baryons considered here belong instead  to the $({\bf 3},{\bf 20_{MS}})$ representation,  where  $\bf 20_{MS}$
is the mixed symmetric
representation of $SU(4)$. For general $N_c$ the spin-flavor representations involve, in the Young tableaux
language,  $N_c$ boxes  and are identified with  the totally symmetric and the
mixed-symmetric representations of type $(N_c-1,1)$  for ground  and excited negative
parity states respectively. Since in the mixed symmetric spin-flavor representation one
box of the Young tableaux is distinguished,  such a box  is associated with  the \lq\lq excited
quark\rq\rq \   in the baryon. In a similar fashion, and without any loss of generality it
is possible to distinguish one box in the ground state multiplet as well.  This is a very
convenient  procedure    that has been used repeatedly in previous works.  The  spin and
isospin quantum numbers of the distinguished box will be denoted with lower cases, and the
corresponding quantum numbers of the rest of $N_c-1$ boxes (which are in a totally symmetric
representation of $SU(4)$  and form the so  called  \lq\lq core\rq\rq ~ of the large $N_c$
baryon) after they are coupled to eigenstates of spin and isospin will be denoted by $S_c$
and $I_c$ respectively.  Notice  that $S_c=I_c$ for totally symmetric representations of
$SU(4)$. For a given core state, the coupling of the excited quark gives eigenstates of
spin $S$ and isospin $I$: $\mid~S,  S_3 ; I , I_3 ; S_c\rangle$.
These states are not in an irreducible representation  of $SU(4)$,
as they are not in an irreducible representation of the permutation group. The totally symmetric
states are given by:
\begin{eqnarray}
\mid S, S_3 ; I , I_3 \rangle_{\mbox{\bf S}} &=&\sum_{\eta=\pm \frac{1}{2}}
C_{\mbox{\bf S}}(S,\eta) \mid S , S_3 ; I , I_3 ; S_c=S+\eta\rangle,
\end{eqnarray}
where
\begin{eqnarray}
C_{\mbox{\bf S}}(S,\pm \frac{1}{2})&=&  \sqrt{\frac{(2 S + 1 \pm 1 ) (N_c+ 1 \mp (2 S+ 1))}{2 N_c ( 2 S + 1)} } \ ,
\end{eqnarray}
while  the mixed symmetric ({\bf MS}) states   $(N_c-1,1)$ are given by:
\begin{eqnarray}
\mid S, S_3,\; I \; I_3\rangle_{\mbox{\bf MS}} &=&\sum_{\eta=\pm \frac{1}{2}}
C_{\mbox{\bf MS}}(I,S,\eta) \mid~S, S_3 ; I , I_3 ; S_c=S+\eta \rangle ,
\end{eqnarray}
where
\begin{eqnarray}
C_{\mbox{\bf MS}}(I,S, \pm \frac{1}{2})&=&\left\{
    \begin{array}{c}
      1~~{\rm if}~~I=S \pm 1   \\
      0~~{\rm if} ~~ I=S \mp 1 \\
      \pm \sqrt{\frac{(2 S + 1 \mp 1 ) (N_c+ 1 \pm (2 S+ 1))}{2 N_c ( 2 S + 1)} } ~~{\rm if} ~~I=S
    \end{array}\right.
\end{eqnarray}
Finally, upon coupling the orbital degrees of freedom, the excited baryons in the
$({\bf 3},{\bf {MS}})$ representation, $\mid~(1,S)J , J_3 ; I , I_3 ; S \rangle_{\mbox{\bf MS}}$,
are obtained.

 Note that there are two sets of $N^*$ states each consisting of two states with the same spin and isospin.
 The physical states are admixtures of such states, and are given by:
\bea
\left(
    \begin{array}{c}
        N^*_{J}  \\
        N^{*'}_{J}
    \end{array}
\right)
&=&
\left(
    \begin{array}{rr}
        \cos{\theta_{2 J }}  & \sin{\theta_{2 J }}  \\
            - \sin{\theta_{2 J }}  &  \cos{\theta_{2 J }}
    \end{array}
\right)
\left(
    \begin{array}{c}
        ^2N^*_{J}  \\
        ^4N^*_{J}
    \end{array}
\right) \ ,
\eea
where $J=\frac{1}{2}$ and $\frac{3}{2}$, $N^{*(')}_J$ are mass eigenstates, and
the two mixing  angles can be constrained to be in the interval $[0 , \pi)$.
Here  the notation $^{2S+1}N^*_{J}$ has been used.

Using the standard definition for the decay width
and averaging over the initial- and summing over
the corresponding final-baryon spins and isospins,
the decay width for each orbital angular
momentum $\ell_P$ and isospin $I_P$ of the pseudoscalar meson
is given by
\begin{equation}
\Gamma^{[\ell_P,I_P]}
= f_{ps} \
\frac{|\sum_q C_q^{[\ell_P, I_P]}\
{\it B}_q(\ell_P,I_P,S,I,J^*,I^*,S^*)|^2}
{\sqrt{(2 J^* + 1)(2 I^*+1)}},
\label{width}
\end{equation}
where the phase space factor $f_{ps}$ is
\begin{equation}
f_{ps} =\frac{k_P^{1+2\ell_P}}{8 \pi^2 \Lambda^{2\ell_P}} \frac{M_{B^*}}{M_B}
\label{fsp}
\end{equation}
Here, we have used that the baryonic operator admits an expansion in $1/N_c$ and has the general form:
\begin{equation}
B^{[\ell_P, I_P]}_{[\mu, \alpha]} =
\left(\frac{k_P}{\Lambda}\right)^{\ell_P}\sum_q \, C_q^{[\ell_P, I_P]}(k_P)
\left( B^{[\ell_P, I_P]}_{[\mu, \alpha]} \right)_q,
\label{exp}
\end{equation}
where
\begin{equation}
\left( B^{[\ell_P, I_P]}_{[\mu, \alpha]} \right)_q = \sum_m
\langle 1, m ; j, j_z  \mid \ell_P, \mu \rangle \
\xi^1_m \ \left( {\it{G}}^{[j, I_P]}_{[j_z,\alpha]} \right)_q,
\end{equation}
and the factor  $\left(\frac{k_P}{\Lambda}\right)^{\ell_P}$ is included to take
into account the chief meson momentum dependence of the partial wave. The scale $\Lambda$ is chosen in what follows to be 200 MeV.
Here, $\xi^1_m$ is an operator that produces a transition from the triplet to the singlet $O(3)$ state,
and $\left( {\it{G}}^{[j, I_P]}_{[j_z,\alpha]} \right)_q$ is a spin-flavor operator that produces the transition from the
mixed-symmetric to the symmetric $SU(4)$ representation. The label  $j$ denotes  the spin of the spin-flavor operator,
and as it is clear, its isospin coincides with the isospin of the emitted meson.
The dynamics of the decay is encoded in the effective dimensionless coefficients
$C_q^{[\ell_P, I_P]}(k_P)$. The reduced matrix elements ${\it B}_q(\ell_P,I_P,S,I,J^*,I^*,S^*)$ appearing
in Eq.(\ref{width})
can be easily calculated in terms of the reduced matrix elements of the spin-flavor operators.
Note that in the present case, where $\ell_P$ can be 0 or 2 only, the
spin-flavor operators can carry spin $j$ that can be 1, 2 or 3.

The terms in the right hand side of Eq.(\ref{exp}) are ordered in powers of $1/N_c$.
As it has been explained in earlier publications\cite{GroundStates1}, the order in $1/N_c$
is determined by the spin-flavor operator. For an $n$-body operator, this order is given by
\begin{equation}
\nu=n-1-\kappa,
\end{equation}
where $\kappa$ is equal to zero for incoherent operators and can
be equal to one or even larger for coherent operators. More details can be found in the following section
where a basis of operators ${\it{G}}$ is explicitly built.

With the definition of effective operators used in this work, all  coefficients
$C_q^{[\ell_P, I_P]}(k_P)$ in Eq.(\ref{exp}) are of zeroth order in $N_c$.
The leading order of the
decay amplitude is in fact $N_c^0$ \cite{GoityMixDecay}.
At this point it is important to comment on the momentum dependence of
the coefficients. The  spin-flavor breakings  in the masses, of both excited
and ground state baryons,  give rise to different values of the momenta $k_P$.
In this work, we have adopted a scheme where the only
momentum dependence assigned to the coefficients is the
explicitly shown factor    $\left(k_P/\Lambda\right)^{\ell_P}$ that
takes into account the chief momentum dependence of the corresponding
partial wave,  and the rest of the dependence is then encoded in the coefficients
of the sub-leading operators.

\section{ Basis of spin-flavor operators}

The spin-flavor operators to be considered in the present paper must
connect a mixed-symmetric with a symmetric representation.
Generators of the spin-flavor group acting on the states
obviously do not produce such connection. However, generators
restricted to act on the excited quark or on the core of
$N_c-1$ quarks can do this. The spin flavor operators can,
therefore,  be represented by products of generators of the
spin-flavor group restricted to act either on the excited
or on the core states.
 In the following the generators acting on the core are denoted by $S_c$,
$G_c$, $T_c$, and the ones acting on the excited quark by $s$, $g$, $t$. The
generators $G_c$  are known to be coherent operators, while all the rest are
incoherent. In order to build a basis of operators for the present problem one
has to consider products of such generators with the appropriate couplings of
spins and isospins. The $n$-bodyness ($n$B) of an operator is given by the number of
such factors, and the level of coherence of the operator is determined by how
many factors $G_c$ appear in the product. It should be noticed that
in the physical case where $N_c=3$, only operators of at most 3B have
to be considered. Still, to order $1/N_c$ there is a rather long list of
operators of given spin $j$ and isospin $I_P$. This list can be drastically
shortened by applying several reduction rules. The first rule is that the
product of two or more generators acting on the excited quark can always be
reduced to the identity operator or to a linear combination of  such  generators.
The second set of rules can be easily
derived for products of operators whose  matrix elements are taken  between a
mixed-symmetric and a symmetric representation. These reduction rules are as
follows (here $\lambda$ represents generators acting on the excited quark and
$\Lambda_c$ represent  generators acting on the core):

\bea
\lambda&=&-\Lambda_c \nonumber\\
(\Lambda_c)_1 (\Lambda_c)_2
&=&- \lambda_1 (\Lambda_c)_2- \lambda_2 (\Lambda_c)_1 + {\rm 1B~ operators}.
\eea
Therefore, only the following types of operators should be considered
\begin{eqnarray}
\qquad \quad \mbox{1B} \qquad & & \qquad \qquad \quad \mbox{2B}
\qquad \quad \qquad \qquad \qquad \quad \mbox{3B} \qquad \nonumber \\
\qquad \lambda \qquad &;&
\qquad \frac{1}{N_c} \ \lambda_1 \ (\Lambda_c)_2 \qquad ;
\qquad \frac{1}{N_c^2} \ \lambda_1 \ (\Lambda_c)_2 \ (\Lambda_c)_3
\label{123B}
\end{eqnarray}

It is convenient to make explicit the transformation
properties of each basic operator under spin $j$ and isospin $t$.
In what follows we use the notation $O^{[j,t]}$ to indicate that the
operator $O$ has spin $j$  and isospin $t$. It is easy to see that
\begin{eqnarray}
\lambda^{[j,t]} &=& s^{[1,0]},\ t^{[0,1]}, \ g^{[1,1]} \nonumber \\
(\Lambda_c)^{[j,t]} &=& (S_c)^{[1,0]},\ (T_c)^{[0,1]},
\ (G_c)^{[1,1]} \label{transbasic}
\end{eqnarray}
For decays in the $\eta-$channels the spin-flavor operators  transform as $[j,0]$ while
for decays in the pion channels they should transform as $[j,1]$, where in both
cases $j=1,2,3$. Knowing the transformation properties of each
basic operator given in Eq.(\ref{transbasic}) it is easy to
construct products of the forms given in  Eqs.(\ref{123B}) with the desired spin and isospin.
Thus, the 1B operators that contribute to a given $[j,t]$ are just
those $\lambda^{[j,t]}$ given in Eq.(\ref{transbasic})
which have the proper spin and isospin quantum numbers. Similarly,
the possible 2B operators are given by the products
$\lambda_1^{[j_1,t_1]}
(\Lambda_c)_2^{[j_2,t_2]}$ coupled to the required
$[j,t]$ by means of the conventional Clebsch-Gordan
coefficients.

In order to construct all possible 3B operators, it is convenient to note
that there are additional reduction rules for the products
$(\Lambda_c)_2 (\Lambda_c)_3$. The starting
point is to consider all possible products of two core operators.
Since one is interested in keeping contributions of at most order
$1/N_c$, and because these products will appear only in  3B
operators, at least one of the operators must be a $G_c$.
Using the reduction relations\cite{DashenManohar},  it
can be shown\cite{GSS05} that the relevant list of independent products of two core
operators turns out to be
\begin{eqnarray}
( \{ T_c, G_c \} )^{[1,2]} \ , \
( [ S_c, G_c ]   )^{[1,1]} \ , \
( \{ S_c, G_c \} )^{[2,1]} \ , \
(G_c \ G_c)^{[2,2]}
\end{eqnarray}
where  $( [ T_c, G_c ] )^{[1,1]} $ denotes $ (T_c G_c)^{[1,1]} - (G_cT_c)^{[1,1]}$, etc..
By coupling any of these operators with one of the excited core
operators $\lambda_1^{[j_1,t_1]}$ (see
Eq.(\ref{transbasic})) to the required $[j,t]$
all the possible 3B operators are obtained.

Using this scheme to couple products
of generators it is straightforward to
construct lists of operators with spin
1, 2 and 3 and isospin 0 and 1.
Further reductions result from the fact that
not all the resulting  operators are linearly independent
up to order $1/N_c$. The determination of  the final set of independent
operators for each particular decay channel is more laborious since it requires the explicitly calculation
of all the relevant matrix elements.
The resulting  basis of independent operators
$\left( O^{[\ell_P, I_P]}_{[m_P,I_{P_3}]}\right)_q$ is shown in Table~1, where for
simplicity the corresponding spin and isospin projections have been omitted.
Finally, the normalized basis operators $\left( B^{[\ell_P, I_P]}_{[m_P,I_{P_3}]}\right)_q$
are defined. They differ from those listed in Table~1 by a normalization
constant which is determined by requiring that, for $N_c=3$, their
largest reduced matrix element should be equal to one for order $N_c^0$ operators
and equal to 1/3 for order $1/N_c$ operators.
\begin{table}[htdp]
\caption{Operator basis}
{\footnotesize
\begin{tabular}{cccccccccc}
\hline
  & $n$-bodyness    &   Name  & Operator & Order in $1/N_c$ \\ \hline
    &  1B   &   $O_1^{[0,1]}$ &
$\left(\xi \ g\right)^{[0,1]}$ &
0 \\[0.5mm] \cline{2-5}
 Pion   &       &  $O_2^{[0,1]}$ &
$\frac{1}{N_c} \ \left(\xi \left(s\ T_c\right)^{[1,1]}\right)^{[0,1]}_{[0,a]}$
& 1 \\[0.5mm]
 S wave   &  2B   &  $O_3^{[0,1]}$ &
$\frac{1}{N_c} \ \left(\xi  \left(t\ S_c\right)^{[1,1]}\right)^{[0,1]}_{[0,a]}$
& 1 \\[0.5mm]
    &       &  $O_4^{[0,1]}$ &
$\frac{1}{N_c} \ \left(\xi \left(g \ S_c\right)^{[1,1]}\right)^{[0,1]}_{[0,a]}$
& 1 \\[0.5mm] \hline
   &   1B  & $O_1^{[2,1]}$ &
$\left(\xi \ g \right)^{[2,1]}_{[i,a]}$
& 0    \\[0.5mm] \cline{2-5}
     &       & $O_2^{[2,1]}$ &
$\frac{1}{N_c} \ \left(\xi \left(s \ T_c\right)^{[1,1]} \right)^{[2,1]}_{[i,a]}$
& $1$ \\[0.5mm]
 Pion    &       & $O_3^{[2,1]}$ &
$\frac{1}{N_c}  \ \left(\xi \left(t \ S_c\right)^{[1,1]} \right)^{[2,1]}_{[i,a]}$
& $1$ \\[0.5mm]
 D wave    &   2B  & $O_4^{[2,1]}$ &
$\frac{1}{N_c} \ \left(\xi \left(g \ S_c\right)^{[1,1]} \right)^{[2,1]}_{[i,a]}$
& $1$ \\[0.5mm]
     &     & $O_5^{[2,1]}$ &
$\frac{1}{N_c} \ \left(\xi  \left( g \ S_c \right)^{[2,1]} \right)^{[2,1]}_{[i,a]}$
& $1$ \\[0.5mm]
     &     & $O_6^{[2,1]}$ &
$\frac{1}{N_c} \ \left(\xi  \left( s \ G_c\right)^{[2,1]} \right)^{[2,1]}_{[i,a]}$
& $0$ \\[0.5mm] \cline{2-5}
     &  3B   & $O_7^{[2,1]}$ &
$ \frac{1}{N_c^2} \ \left(\xi
\left( s \left( \left\{ S_c, G_c \right\} \right)^{[2,1]}
\right)^{[2,1]}\right)^{[2,1]}_{[i,a]}$
& $1$ \\[0.5mm]
     &     & $O_8^{[2,1]}$ &
$\frac{1}{N_c^2} \ \left(\xi
\left( s \left( \left\{ S_c, G_c \right\} \right)^{[2,1]}
\right)^{[3,1]}\right)^{[2,1]}_{[i,a]}$
& $1$ \\[0.5mm] \hline
 Eta    &    1B  &  $O_1^{[0,0]}$ &
$\left( \xi \ s \right)^{[0,0]}_{[0,0]}$
& 0 \\[0.5mm] \cline{2-5}
S wave  &    2B  &  $O_2^{[0,0]}$ &
$\frac{1}{N_c} \left(\xi \left(s \ S_c\right)^{[1,0]} \right)^{[0,0]}_{[0,0]}$
& 1 \\[0.5mm] \hline
 Eta       &  1B   & $O_1^{[2,0]}$ &
$\left( \xi \  s \right)^{[2,0]}_{[i,0]}$
&  0 \\[0.5mm] \cline{2-5}
 D wave    &  2B   & $O_2^{[2,0]}$ &
$\frac{1}{N_c} \left(\xi \left(s \ S_c\right)^{[1,0]} \right)^{[2,0]}_{[i,0]}$
& 1  \\[0.5mm]
        &       & $O_3^{[2,0]}$ &
$\frac{1}{N_c} \left(\xi \left(s\ S_c\right)^{[2,0]} \right)^{[2,0]}_{[i,0]}$
& 1\\[0.5mm] \hline
\end{tabular}
}
\end{table}

\section{Results}
The  different  empirical S- and D-wave decay widths used in the analysis are the ones provided by the
PDG\cite{PDG}.  The values for the widths and branching ratios are taken as the ones
indicated there as \lq\lq our estimate\rq\rq, while the errors are determined from the corresponding ranges.
The corresponding partial widths obtained from those values are explicitly displayed in Table~3. The entries
indicated  as unknown reflect channels for which no width is provided by the PDG or where the
authors consider that the input is unreliable, such as in the  $\pi \Delta$ decay modes of the $N(1700)$ and
$N(1675)$ and the D-wave $\eta-$ decay modes. At this point it is important to stress the marginal precision
of the data for the purposes of this work. This work performs an analysis at order $1/N_c$, which means that
the theoretical error  is order $1/N_c^2$.  This implies that amplitudes are affected by a theoretical
uncertainty at the level of 10\%. Thus, in order  to pin down the coefficients of the subleading operators,
the widths provided by the data should not  be affected by errors larger than about 20\%. As shown in Table~3,
the experimental errors are in most entries 30\% or larger. In consequence, the determination of the
subleading effective coefficients is affected by large errors as the results below show.

Before presenting the results of the fits,  it is convenient to derive some parameter
independent relations that can be obtained to leading order.  These relations serve as a test
of the leading order approximation.  Since at this order there are only four coefficients and two angles to be
fitted, and there are a total of twenty partial widths (excluding all D-wave $\eta$ channels
but including kinematically forbidden $\eta$-channel decays), there are fourteen  independent
parameter free relations that can be derived. These relations are more conveniently written in
terms of reduced widths, i.e., widths where the phase space factor $f_{sp}$ (see Eq.(\ref{fsp}))
has been removed and denoted
here by $\tilde \Gamma$. Considering the S-wave decays in the $\pi$ mode, there are six decays
and three parameters in the fit. Thus, three parameter free relations must follow.
These relations and the corresponding comparison with experimental values read:
\begin{equation}
\left.
\begin{array}{cccccccc}
 &{\begin{array}{c}\tilde{\Gamma}_{N(1535)\to\pi N}\\
 +\tilde{\Gamma}_{N(1650)\to\pi N}
 \end{array} }&:&{\begin{array}{c} \tilde{\Gamma}_{N(1520)\to\pi \Delta}\\+
 \tilde{\Gamma}_{N(1700)\to \pi \Delta} \end{array} }
 &:&\tilde{\Gamma}_{\Delta(1620)\to\pi N}&:&\tilde{\Gamma}_{\Delta(1700)\to\pi\Delta}\\
{\rm Th.}& 1        &:& 1                            &:&0.17                 &:& 0.42 \\
{\rm Exp.}&       1 &:& {\rm unknown}     &:&0.19\pm 0.07 &:& 0.62\pm 0.33
\end{array}
\right.
\nonumber
\end{equation}
Within the experimental errors the relations are satisfied.
In a similar fashion, relations
involving D- wave decays in the $\pi$ mode  can be obtained. There are in this case four parameters to fit and
eleven partial widths. Thus, seven parameter free relations can be obtained. In general these relations
are quadratic and/or involve some of the unknown decay widths. However, three testable linear relations
can be obtained. The simplest of them reads,
\begin{eqnarray}
2\ \tilde{\Gamma}_{\Delta(1620)\to\pi \Delta} + \tilde{\Gamma}_{\Delta(1700)\to\pi \Delta}&=&
8\ \tilde{\Gamma}_{\Delta(1700)\to\pi N} + \frac{15}{4}\ \tilde{\Gamma}_{N(1675)\to\pi N} \nonumber \\
{\rm Exp.}  \qquad \qquad 5.24 \pm 1.95 \qquad \qquad &=& \qquad \qquad 2.19 \pm 0.62
\nonumber
\end{eqnarray}
As we see this relation is not well satisfied
by the empirical data. Something similar happens with the other two.
Thus, one can anticipate a poor leading order  fit to the D-wave pion decays.
\begin{table}[htdp]
\caption{ Fit parameters. Fit {\#1}: Pion S-waves LO.
In this case there is a four-fold ambiguity for
the angles $\{ \theta_1 , \theta_3 \}$ given by the two values
shown for each angle. Fit {\#2}: Pion
S and D-waves, eta S-waves, LO. In this case there is a two-fold ambiguity for
the angle $\theta_1$. For the angle $\theta_3$ there is an {\it almost} two-fold ambiguity
given by the two values indicated in parenthesis and which only differ in the two slightly
different values of $C^{[2,1]}_6 $. Fit {\#3}: Pion S and D-waves, eta S-waves,
NLO, no 3-body operators. No degeneracy in $\theta_1$  but
{\it almost} two-fold ambiguity in $\theta_3$ given by the two values indicated in parenthesis.
Values of coefficients which differ in the corresponding fits are indicated in parenthesis.}
{\footnotesize
\begin{tabular}{cccc}
\multicolumn{4}{c}{ } \\
\hline
Coefficient      &   \#1 LO     &   \#2 LO     &     \#3 NLO     \\  \hline
$C^{[0,1]}_1 $   &  $31 \pm 3 $ &   $31 \pm 3$           &   $  23\pm 3  $ \\
$C^{[0,1]}_2 $   &    -         &      -                 &   $ (7.4,32.5)   \pm (27,41) $ \\
$C^{[0,1]}_3 $   &    -         &      -                 &   $ (20.7,26.8)  \pm (12,14) $ \\
$C^{[0,1]}_4 $   &    -         &      -                 &   $ (-26.3,-66.8)\pm (39,65) $ \\  \hline
$C^{[2,1]}_1 $   &    -         & $4.6\pm 0.5$           &   $ 3.4\pm 0.3$ \\
$C^{[2,1]}_2 $   &    -         &      -                 &   $-4.5\pm 2.4$ \\
$C^{[2,1]}_3 $   &    -         &      -                 &   $ (-0.01,0.08)\pm 2$ \\
$C^{[2,1]}_4 $   &    -         &      -                 &   $ 5.7\pm 4.0$ \\
$C^{[2,1]}_5 $   &    -         &      -                 &   $ 3.0\pm 2.2$ \\
$C^{[2,1]}_6 $   &    -         & $(-1.86,-2.25)\pm 0.4$ &   $ -1.73\pm 0.26$ \\
$C^{[2,1]}_7 $   &    -         &      -                 &     -           \\
$C^{[2,1]}_8 $   &    -         &      -                 &     -           \\ \hline
$C^{[0,0]}_1 $   &    -         &  $11\pm 4$             &    $17\pm 4$           \\
$C^{[0,0]}_2 $   &    -         &      -                 &     -           \\ \hline
$\theta_1    $
                 & $ \begin{array} {c}1.62 \pm 0.12  \\ 0.29 \pm 0.11 \end{array} $
                 & $ \begin{array} {c}1.56 \pm 0.15  \\ 0.35 \pm 0.14 \end{array} $
                 & $ \begin{array} {c} 0.39 \pm 0.11 \end{array} $
                   \\ \hline
$\theta_3    $   & $ \begin{array} {c}3.01 \pm 0.07  \\ 2.44 \pm 0.06 \end{array} $
                 & $ \begin{array} {c}(3.00,2.44) \pm 0.07         \end{array} $
                 & $ \begin{array} {c}(2.82,2.38) \pm 0.11         \end{array} $
                   \\
\hline
$\chi^2_{\rm dof} $ &   0.25     &  1.5  &       0.9                   \\
${\rm dof}$         &     2     &  10   &       3                       \\ \hline
\end{tabular}
\label{defaulta}
}
\end{table}

In Table~2 the results of several fits are displayed. In these fits the decay amplitudes are expanded
keeping  only the terms that correspond to the order in $1/N_c$ of the fit. Similarly, when performing the
fits the  errors have been taken to be equal or larger than the expected accuracy of the fit ($30 \%$ to the
LO fits and about $10\%$ for the NLO ones).

The first LO fit only
considers the S-wave $\pi$-modes. One notices here that both $\theta_1$ and $\theta_3$ have a
two fold ambiguity at this order. The second leading order fit includes the D-waves and $\eta$-modes. The
angles remain within errors equal to the ones from the first fit.  Table~3 shows that the  $N(1535)\to
\eta N$  width results to be a factor four smaller than the empirical one (this having, however, a rather
generous error). In the D-waves, several widths involving decays with a $\Delta$ in the final state are also
too small. The S-wave $\pi$-modes are well fitted and there is no real need for NLO  improvement. In the
D-wave decays in the $\pi$ channel there are two leading order  operators that contribute.  The fit shows
that the 1B operator has a coefficient whose magnitude  is a factor two  to three larger than that of the
coefficient of the 2B operator.  The 1B D-wave  operator  as well as  the 1B S-wave operator $O^{[0,1]}_1$
stem from   the 1B coupling of the pion via the axial current. Such a coupling naturally occurs as the
dominant coupling in the chiral quark model\cite{Manohar:1983md}.

\begin{table}[htdp]
\caption{ Partial widths resulting from the different fits in Table~\ref{defaulta}.
Values indicated in parenthesis correspond to the cases in which {\it almost} degenerate values of
$\theta_3$ lead to different partial widths.}
{\footnotesize
\begin{tabular}{cccccc}
\multicolumn{6}{c}{ } \\
\hline
 \multicolumn{2}{c}{Decay} &\hspace{.2cm} Emp.    \hspace{.2cm}   &
 \hspace{.1cm} \#1 LO  \hspace{.1cm}   &   \hspace{.1cm} \#2 LO  \hspace{.1cm}
 &   \hspace{.1cm}  \#3 NLO   \hspace{.1cm}  \\
 & & [MeV] & [MeV] & [MeV] & [MeV] \\ \hline

  & $N(1535)\rightarrow \pi N$      & $68 \pm 27$  &
  74 &         62              &      (58,68)        \\

  & $N(1520)\rightarrow \pi \Delta$ & $10 \pm 4$   &
  10 &      9.7                 &       9.5       \\
\hspace{.5cm}$\pi$ \hspace{.5cm}
  & $N(1650)\rightarrow \pi N$      & $121 \pm 40$ &
132  &     144                   &       122       \\
S-wave
  & $N(1700)\rightarrow \pi \Delta$ & unknown      &
 175 &     175                   &     (259,156)         \\

  & $\Delta(1620)\rightarrow \pi N$ &  $38 \pm 13$     &
  35 &      35                  &      38       \\

  & $\Delta(1700)\rightarrow \pi \Delta$ &  $112 \pm 53$     &
  81 &        81                &     (135,112)        \\
\hline

  & $N(1535)\rightarrow \pi \Delta$      & $1\pm 1$  &
     &           .01             &     0.5         \\

  & $N(1520)\rightarrow \pi    N $       & $67\pm 9$  &
     &          70              &       65       \\

  & $N(1520)\rightarrow \pi \Delta$      & $15\pm 3$  &
     &          2.8              &     13         \\

  & $N(1650)\rightarrow \pi \Delta$      & $7 \pm 5$  &
     &          0.12              &      8        \\
$\pi$
  & $N(1700)\rightarrow \pi N$           & $10 \pm 7$  &
     &         10               &      (11,9)        \\
D-wave
  & $N(1700)\rightarrow \pi \Delta$      & unknown  &
     &         4               &     (4,9)         \\

  & $N(1675)\rightarrow \pi N$           & $72 \pm 12$  &
     &         85               &      76        \\

  & $N(1675)\rightarrow \pi \Delta$      & unknown  &
     &        45                &      79        \\

  & $\Delta(1620)\rightarrow \pi \Delta$      & $68 \pm 26$  &
     &        30                &      87        \\

  & $\Delta(1700)\rightarrow \pi N$           & $45 \pm 21$  &
     &        49                &     32         \\


  & $\Delta(1700)\rightarrow \pi \Delta$      & $12 \pm 10$  &
     &        15                &      18        \\
\hline
$\eta$
  & $N(1535)\rightarrow \eta N$              & $64\pm 28$  &
     &        17                &     (57,61)         \\
S-wave
  & $N(1650)\rightarrow \eta N$              & $11\pm 6$  &
     &        14                &      12        \\ \hline
\end{tabular}
\label{resul}
}\end{table}
   The NLO fit involves  a rather large number of effective constants. Since there are only sixteen  data available,
  some operators must be discarded for the purpose of the fit. It is reasonable to choose to neglect
  the 3B operators and the subleading S-wave operator for $\eta$ emission. The NLO fit has been carried
  out by demanding that the LO coefficients are not  vastly different from their values obtained in the LO fit.
  This demand is reasonable if the assumption is made that the $1/N_c$ expansion makes sense. The fact that the
  NLO coefficients do not have unnaturally large values with respect to the scale set by the LO fit indicates the consistency of the assumption. This
  clearly is no proof, however, that the $1/N_c$ expansion is working. As mentioned earlier, the chief
  limitation here is due to the magnitude of the errors in the inputs.   This leads to results for the NLO
  coefficients being affected with rather large uncertainties. Indeed, no clear NLO effects can be pinned down,
  as most NLO coefficients are no more than one standard deviation from zero.  Because  the number of
  coefficients is approximately equal to the number of inputs, there are important correlations between  them.
  For instance, the S-wave NLO coefficients are very correlated with each other and  with  the angle $\theta_3$.
  For the S-waves the LO fit is already excellent, and therefore nothing significantly  new is obtained by including
  the NLO corrections.
  Correlations are smaller for the D-wave coefficients. As mentioned before, here the LO fit has room for
  improvement, and thus the NLO results are more significant than in the case of the S-waves. Still, no clear
  pattern concerning the NLO corrections is observed. One interesting point, however,  is that without any significant
  change in the value of $\theta_1$ the $\eta$-modes are now well described. The reason for this is that in the
  LO fit the matrix elements of the operator $O_1^{[0,0]}$ were taken to zeroth order in $1/N_c$, while in the
  NLO fit the $1/N_c$ terms are included. These subleading corrections enhance the amplitude for the
  $^2N^*_{1/2}$ and suppress the amplitude for the $^4N^*_{1/2}$. This along with an increment in the coefficient
  brings the fit in line with the empirical widths. One important point is that at NLO the two fold ambiguity in
  $\theta_1$ that results at LO is eliminated. The smaller mixing angle turns out to be selected. The angle
  $\theta_3$ remains ambiguous and close to the values obtained in the LO fits. It should be noticed
  that the present values of both mixing angles are somewhat different
  from the values $\theta_1=0.61, \ \theta_3=3.04$  obtained in other analyses\cite{CaroneGeorgi,CapstickRoberts,lll}.

\section{Conclusions}

In this work we have analyzed the strong decays of the negative parity excited baryons in the $1/N_c$ expansion
assuming that configuration mixings can be neglected. A basis of effective operators, in which the S- and D-wave
amplitudes are expanded, was constructed to order $1/N_c$. All dynamical effects are then encoded in the effective
coefficients that enter in that expansion. We find that a consistent description of the decays within the $1/N_c$
expansion is possible. Indeed, up to the relatively poor determination of $1/N_c$
corrections that results from the magnitude of the errors in the
input widths, these corrections are of natural size. A few clear
cut observations can be made. The most important one is that the
S-wave $\pi$- and $\eta$-channels are well described by the
leading order operators (one for each channel) provided one
includes  the contributions subleading in
$1/N_c$ in the matrix elements for the $\eta$-decays. The mixing angle $\theta_1$ is then determined by these
channels up to a twofold ambiguity, which is lifted when all
channels are analyzed at NLO. The angle $\theta_3$ is also
determined up to a two fold ambiguity at LO. The ambiguity remains
when the NLO is considered.   The subleading
operators are shown to be relevant to fine tune the S-wave decays
and improve the D-wave decays. Because of the rather large error
bars in and  significant correlations  between the resulting
effective coefficients, no clear conclusions about the physics
driving the $1/N_c$ corrections can be made.  The mixing angles
$\theta_1 = 0.39 \pm 0.11$ and $\theta_3 = (2.82, 2.38) \pm 0.11$
that result at NLO are similar to the ones determined at LO. They are ,
however,  somewhat different from the angles $\theta_1=0.61$ and
$\theta_3=3.04$ obtained in other analyses\cite{CaroneGeorgi,CapstickRoberts,lll}.

\section*{Acknowledgments}

The material presented here is based on work done in collaboration
with Jos\'e Goity and Carlos Schat. I would like to thank them as well
as Rich Lebed and Toni Pich for the help in putting together this
``Large $N_c$ QCD" workshop.  I also appreciate the fine assistance of the
ECT$^*$ staff. This work was partially supported by CONICET (Argentina)
grant \# PIP 02368 and by ANPCyT (Argentina) grant \# PICT 00-03-08580.

\end{document}